\documentclass[a4paper,10pt]{article}
\usepackage{cite}
\usepackage{enumitem}
\usepackage{graphicx}
\usepackage{subfigure}
\usepackage{amssymb}
\usepackage{amsmath}
\usepackage{bbm}
\usepackage{color}

\title{Testing isotropy of Cosmos with WMAP and PLANCK data}
\author{Pranati K. Rath$^1$\footnote{email:pranati@iitk.ac.in} and Pramoda Kumar Samal$^2$\footnote{email:pksamal@iopb.res.in} }
\date{}

\begin{document}
\maketitle
\begin{center}
{$^{1}$Department of Physics, Indian Institute of Technology Kanpur, Kanpur, 208016, India \\ }
{$^2$Department of Physics, Utkal University, Bhubaneswar, 751004, India\\}
\end{center}
 
\begin{abstract}
  
In recent years, there have been a large number of studies which support violation of statistical isotropy. 
Meanwhile there are some studies which also found inconsistency. We use the power tensor method defined earlier 
in the literature to study the new CMBR data. The orientation of these three orthogonal vectors, 
as well as the power associated with each vector, contains information about possible violation of statistical isotropy.  
This information is encoded in two entropy measures, the power-entropy and alignment-entropy.  
We apply this method to WMAP 9-year and PLANCK data. 
Here, we also revisit the statistics to test high-$l$ anomaly reported in our earlier paper and find that 
the high degree of isotropy seen in earlier WMAP 5-year data is absent in the revised WMAP-9 year data.
\end{abstract}


\section{Introduction}
The existence of a preferred axis from various data particularly polarization of radio waves, 
optical polarization from quasars, cluster peculiar velocity and also low multipoles in CMB $l =1,2,3$ 
which are pointing towards Virgo is supported by many observations \cite{Birch1982,Jain1999,Hutsemekers1998,Costa2004,Schwarz2004}.
The low-$l$ anomalies are tested using various statistics \cite{Copi2004,Copi2007}.
Here, we used the power tensor technique \cite{Ralston2004,Samal2008,Samal2009} to study the 
the low-$l$ anomaly such as alignment between CMB quadrupole and octopole in PLANCK data in comparison to WMAP 9-year data. 
Here, we revisit the statistics defined earlier in \cite{Samal2008,Samal2009} to study 
the low-$l$ alignment as well as the high-$l$ anomaly seen in \cite{Samal2009} for WMAP-9 year foreground cleaned data.

The measured temperature ansotropy of CMB can be expanded in spherical harmonics as
\begin{equation}
\Delta T(\hat n) = \sum_{lm} a_{lm} Y_{lm}(\hat n)\,.
\end{equation} 
Statistical isotropy implies,
\begin{equation}
\langle{a_{lm}a^*_{l'm'}}\rangle = C_l \delta_{ll'}\delta_{m,m'}\,.
\end{equation} 

The second rank power tensor that can be formed from the products of the spherical harmonics
coefficients $a_{lm}$ is defined as
\begin{equation}
A_{ij}(l) = {1\over l(l+1)} \sum_{m,m'} a^*_{lm}(J_iJ_j)_{mm'}a_{lm'}\,,
\end{equation}
where $J_i$ ($i=1,2,3$) are the angular momentum operators in
representation $l$. The ensemble average of the power tensor is 
\begin{equation}
\langle{A_{ij}(l)}\rangle = \frac{C_l}{3}\delta_{ij}
\end{equation} 
For each multipole $l \ge 2 $, there exist three rotationally invariant 
eigenvalues of $A_{ij}(l)$. The sum of these three eigenvalues gives the 
usual power $C_l$. The isotropic CMB data predicts all three equal eigenvalues should be degenrate and equal to $C_l/3$.
Taking into account the eigenvalues of the power tensor, the
power entropy \cite{Samal2009} is defined as,
\begin{equation}
 S = -\sum_{i} (\tilde \lambda^{i}) \log(\tilde \lambda^{i})
\end{equation}
where $i = 1,2,3$ corresponds to three eigenvalues of the power tensor matrix. 
The normalized eigenvalues of the power tensor are given as
\begin{equation}
 \tilde \lambda^{i} = \frac{\lambda^{i}}{\sum_{i}\lambda^{i}}
\end{equation}
The range of power entropy $S$ is $0 \le S \le \log(3)$.
The power entropy has maximum value $\log(3)$ for isotropic CMB prediction, 
where all three eigenvalues are equal. Any excess of power along one of the 
eigenvector of power tensor results in low power entropy 
compared to isotropic prediction. Since the largest eigenvalue makes the 
largest contribution to power spectrum, we single
out the eigenvector corresponding to the largest eigenvalue and call it as the
principal eigenvector (PEV).
If the data is anisotropic, a large number of vectors may lie along one direction 
or in preferred plane. Here, the sign of the eigenvectors is meaningless. 
To study the alignment of vectors over a range of multipoles,
the alignment matrix is defined as
\begin{equation}
 X_{ij}(lmax) = \sum_{l = 2}^{lmax}\tilde e_{i}^l\tilde e_{j}^l \,,
\end{equation}
where $\tilde e_{i}, i = 1,2,3$ are the PEVs of power tensor. The eigenvalues of the
alignment matrix $X$ are a probe of the shape of the bundle collected from 
$2\le l\le l_{max}$. The anisotropy in the data can be probed by computing 
the alignment entropy $S_X$ using the alignment matrix $X$.
The alignment entropy $S_X$ is defined as
\begin{equation}
 S_X = -tr({\tilde \rho}_X \log(\tilde \rho_X)) \,,
\label{algentro}
\end{equation}
where $\tilde \rho_X = \frac{X}{tr(X)}$.
For isotropic CMB, $S_X \sim \log(3)$. The alignment entropy is
independent of the power entropy. A very low value of $S_X$ compared to $\log(3)$ indicates a violation of isotropy.
Here we mainly focus on the low-$l$ anomaly such as alignment between 
CMB quadrupole and octopole in PLANCK data in comparison to WMAP 9-year data 
and also the high-$l$ anomaly seen in \cite{Samal2009} using the
foreground cleaned maps provided by WMAP team.

\section{Statistics}
The Significance of anisotropy is detremined by comparing the result for real data with that corresponding 
to 10,000 isotropic randomly generated CMBR data. The significance is quoted in terms of the P-value, which is 
defined as the probability that a random isotropic CMB map may yield a statistic larger than that seen in data. 
We set  preliminary level of
statistical significance using $P-values$ $0.05$. 
The significance of $P-values$
is calculated using the binomial distribution. 
So the probability to encounter $k$ instances of
passing defined by probability $p$ in $n$ trials is
\begin{equation}
P_{bin}(k, \,p, \,n)= p^k(1-p)^{(n-k)}n!/(n-k)!k!.
\label{eqnbin}
\end{equation}
In assessing many P-values, we find the cumulative probability as calculated in \cite{Samal2009}.
We calculate the cumulative binomial probabilities as
\begin{equation}
 P_{bin}(k \geq k_* , \, p, \,n)= \sum^n_{k =k_*} \, P_{bin}(k, \,p, \,n).
\label{eqncumbin}
\end{equation}
Both the probability and cumulative probability tell us how the observed data support the
isotropic random realizations.
\section{CMB data}
We use WMAP 9-year ILC map \cite{Bennett2013} henceforth known as WILC9 and PLANCK's
NILC, SMICA and SEVEM CMBR maps \cite{Planck2014b}.
Here we restrict ourselves to the multipole range $2 \le l \le 50$, since the statistical and
the systematic errors lie within the cosmic variance in this range.
For WILC9, we use the $KQ85$ mask and in the case of PLANCK maps, we use the
CMB-union mask ($U73$) to eliminate the contribution from the galactic foregrounds. We filled the
masked region by simulated isotropic CMB data with appropriate noise for WILC9. While in the case of
PLANCK data, we do not consider the contribution of noise as it is too bulky.
Since the filling of the masked region with random realization produces different
result for different realizations for the full sky \textquotedblleft{data}\textquotedblright map, we obtain the final results
by taking their average over $100$ such filled data maps for both WILC9 and SMICA maps.

At large-$l$, the WILC9 map is not reliable. Hence, we use the individual foreground cleaned Differencing Assembly (DA) maps,
$Q1$, $Q2$, $V1$, $V2$, $W1$, $W2$, $W3$, $W4$, provided by the WMAP team to study the anisotropies in the large multipoles.
We divide the high multipole region into three regions as $2 \leq l \leq 300$,
$150 \leq l \leq 300$ and  $250 \leq l \leq 300$.
We compute the P-values from random realizations including the appropriate detector noise for each band.
\section{Result}
\subsection{Alignment of Quadrupole and octopole}
We test the alignment for WILC9 
and PLANCK maps (NILC, SMICA, SEVEM) using the power tensor
technique \cite{Ralston2004,Samal2008,Samal2009}. We compute the PEVs for both
quadrupole ($l=2$) and octopole ($l=3$) and extract the angle between these two vectors. The PEVs of quadrupole and octopole
for different maps and the corresponding angle between them are given in the Table \ref{tb:PEV}.
The probability of the distribution for two different axes $\hat n$ and 
$\hat n'$ to align within an angle $\theta$ whcich is given as
\begin{equation}
 P(\cos \theta)= (1-\cos\theta)
\end{equation}
where $\cos\theta =|\hat n\cdot \hat n'|$.
The probability of alignment also quoted in Table \ref{tb:PEV}.

\begin{table}[h!]
\centering
\begin{tabular}{|c|c|c|c|c|}
\hline
                  &  $l=2$	           & $l=3$ 		 & $\theta_{23}$ & $1-cos\theta_{23}$  \cr
\hline
WILC9 & $(0.248,0.435,-0.865)$ & $(0.240,0.387,-0.890)$ & $3.87^o$  & $0.00185$ \cr
\hline
NILC  & $(0.087,0.238,-0.967)$ 	  & $(0.240,0.399,-0.884)$ & $13.91^o$ & $0.028$ \cr
\hline 
SMICA & $(0.144,0.347,-0.926)$    & $(0.284,0.405,-0.868)$ & $9.87^o$  & $0.013$  \cr
\hline
SEVEM & $(0.248,0.435,-0.865)$    & $(0.240,0.386,-0.890)$ & $4.22^o$  & $0.00185$ \cr
\hline
\end{tabular}
\caption{Alignments of PEVs for $l = 2$ and $l = 3$ for various maps and the corresponding angle between them.}
\label{tb:PEV}
\end{table}
The alignment of the eigenvectors in the WILC9 and PLANCK (SMICA, NILC and SEVEM) at low-$l$
may be a signal of a fundamental anisotropy or could also be caused by foreground contamination. 
However, the possibility of the foregrounds for the observed 
alignment in CMBR data has been ruled out in Ref.\cite{Aluri2011} and hence it is most likely cosmological in nature.
The alignment is better for WILC9 with $\theta_{23}=3.87^o$ 
and the PLANCK SEVEM map with $\theta_{23}=4.22^o$. But the alignment angle is quite large for SMICA and NILC map.
Hence low-$l$ alignment purely depends upon the cleaning pipeline used to obtain the foreground 
cleaned map.
\subsection{Axial alignments}
\subsubsection{Region $2\leq l \leq11$}
The signals obtained from the region $2\leq l \leq11$ are highly significant as observed in Ref.
\cite{Copi2007}. So it is interesting to analyze this multipole range using both WILC9 and PLANCK data. 
We measure the alignment between multipoles 
by comparing the PEVs of the power tensor matrix.
There are $3$ PEVs corresponding to multipole $l = 2,3,9$ for 
WMAP 5-year ILC map as seen by Ref.\cite{Samal2008}.
We find that there are $2$ PEVs with multipole 
$l = 2,3$ for WILC9, $3$ PEVs with $l = 2,3,9$ for NILC, $2$ PEVs with 
$l = 2,3$ for SMICA, $2$ PEVs with $l = 2,3$ for SEVEM maps which shows alignment with quadrupole. 
So WILC9, SMICA and SEVEM maps no longer support the earlier observation \cite{Copi2007,Samal2008}. 
The only map which support the earlier observation is NILC. 
Hence, we find that WILC9 and the PLANCK data does not support the strong signal of alignment over the multipole range $2\leq l \leq11$.

\subsubsection{Region $2\leq l \leq 50$}
As reported earlier in Ref.\cite{Samal2008}, the WMAP 5-year ILC map also shows alignment with the
quadrupole axis in the low multipole range $2 \leq l \leq 50$ with high 
significance. For 5-year map, there are $6$ multipoles with $l=3,9,16,21,40,43$ aligned with 
quadrupole. Hence, the alignment with the quadrupole is seen over a relatively 
large range of $l$ values.
We also test for the alignment of multipoles in this multipole range for WILC9 and PLANCK maps. 
The list of the multipoles whose P-value of alignment is less than $5\%$ 
with quadrupole is given in Table \ref{tb:Ptable1}. There are $4$, $5$, $5$ and $5$ multipoles
which shows the alignment with the quadrupole for WILC9, NILC, SMICA, 
and SEVEM maps respectively. In the WILC9, $l=9, 40$ multipoles are no longer
aligned with the quadrupole as seen in the WMAP 5 year data.
So there is a very mild signal of anisotropy in this range $2 \le l \le 50$ of multipole. 
The net significance of alignment with quadrupole is computed using binomial probability and cumulative
binomial probability using Eq. (\ref{eqnbin}) and Eq. (\ref{eqncumbin}) is given in Table \ref{tb:cumpro2}. 
 \begin{table}[h!]
   \begin{tabular}{|c|c|c|c|}  
\hline 
WILC9 & NILC & SMICA & SEVEM  \cr 
\hline 
  3	&3  	 &3	&3	  \cr 
 16	&9  	 &16	&16	\cr
 21	&16 	 &28	&28	\cr
 44	&28 	 &40	&40	 \cr
  	&40      &46	&44 	\cr
\hline 
 \end{tabular}
  \caption{List of multipoles whose P-values of coincidence with quadrupole is less than $5\%$ for $l \leq 50$.}
  \label{tb:Ptable1}  
\end{table}

\begin{table}[h]
\begin{tabular}{|c|c|c|c|c|}
\hline
&WILC9 & NILC  & SMICA  & SEVEM \cr
\hline
Pro 	& $0.12$    & $0.06$   & $0.05$      & $ 0.03$ \cr
\hline
CumPro  &  $0.20$   & $ 0.09$ & $ 0.07$     & $0.04$  \cr
\hline  
\end{tabular}
\caption{Net significance of observing $P \le 0.05$ for the multipoles that are well aligned with 
quadrupole $l = 2$ in the multipole range $2 \le l \le 50$. }
\label{tb:cumpro2}
\end{table}
\section{Anomaly in high $l$}
We next test for alignment of PEVs over a large range of multipoles.
Our motivation is to verify whether the anisotropy found in low-$l$ multipole region 
continues to hold for a larger range of multipoles or are there any additional anomalies present in the data.
In Ref.\cite{Samal2009} it was found that the WMAP foreground cleaned 3-year and 5-year data in $W$ band for 
high multipoles shows unusual isotropy.
The alignment entropy was so large that the probability to obtain
this from a random samples exceeds $99.99\%$ in the multipole range $150 \leq l \leq 300$ 
and almost the same probabilities are found for the multipole range $2 \leq l \leq 300$ and 
$250 \leq l \leq 300$. In contrast, the 
$Q$ band shows the signal of anisotropy with probability less than $0.01\%$. Since the 
$Q$ band is highly contaminated, one might assume that the anisotropy found in the 
$Q$ band would be due to the foreground contamination. 
We find that the WMAP foreground cleaned 9-year 
$Q$ band data also shows signal of anisotropy with same probability.

The high level of isotropy of $W$ band goes away when the  
authors lowered the level of noise in the simulated random realizations.~\cite{Samal2009}
Noise maps are generated by multiplying $\frac{\sigma_0}{\sqrt{N_p}}$, where $\sigma_0$ is the
noise per observation and $N_p$ is the effective number of observation at each pixel, with a 
Gaussian distribution having zero mean. In generating the noise map they have used $\sigma_0$ value 
as $5.883 (W_1),\ 6.532 (W_2),\ 6.885 (W_3)$ and $6.744 (W4)$.  
Reducing the value of $\sigma_0$ by two units they found that the P-value decreases to $92$\%.
 However such a large change in the value of $\sigma_0$ is not acceptable. 
To address this issue, we analyze these range of multipoles for WMAP 9-year 
foreground cleaned data. We find that the WMAP 9-year foreground cleaned data does not show such a high level of isotropy. 
The alignment entropy $S_X$ and the corresponding P-values 
for different range of multipoles are given in the Table \ref{tb:frgnd_iso}. The P-value is less than $50\%$ except the 
$W1$ band in the multipole range $2 \leq l \leq 300$ and in $150 \leq l \leq 300$. For the 
multipole range $250 \leq l \leq 300$, the P-value for $W1$ band is also less than $50\%$.
Hence the WMAP 9-year data does not suffer from the anomaly seen in the $W1$ band in the 3-year and 5-year data. 
\begin{table}[h!]
\centering
\begin{tabular}{|c|c|c|c|c|}
\hline
& $W1$  & $W2$  & $W3$  & $W4$ \cr
\hline
$S_X(2, 300 )$ & 1.0934  & 1.0925 & 1.0945 & 1.0948  \cr
$P(\%) $ & 55 & 44 & 18 & 24 \cr
\hline
$S_X(150, 300 )$  & 1.0817 & 1.0838 & 1.0889 & 1.090  \cr
$P(\%) $ & 74 & 37 & 12 & 17  \cr
\hline
$S_X(250, 300 )$  & 1.0657 & 1.0570 & 1.0736 & 1.0827  \cr 
$P(\%) $ & 45 & 34 & 9 & 6 \cr
\hline
\end{tabular}
\caption{ Alignment entropy $S_{X}$ and corresponding P-values (in \%)
for WMAP 9 year foreground cleaned temperature DA maps over the three multipole ranges.}
\label{tb:frgnd_iso}
\end{table}

\section{Conclusion}
We tested for the statistical low-$l$ alignments of the WILC9 and PLANCK 
CMBR data by extracting three orthogonal eigenvectors and the corresponding 
eigenvalues for each multipole $l$. We also measure the dispersion in the 
eigenvalues with the help of power entropy which provides
the measure of the statistical isotropy. We study the
dispersion in the PEV by constructing 
alignment matrix for a range of multipoles.
In the multipole range $2 \le l \le 11$, we do not find any strong signal of anisotropy 
as seen in Refs.~\cite{Copi2007} and\cite{Samal2008}. 
In the multipole range $2 \le l \le 50$, we find $4$, $5$, $5$ and $5$ multipoles
which are aligned with the quadrupole for WILC9, NILC, SMICA,
and SEVEM maps respectively. So there is very mild signal of anisotropy.
The issue of highly isotropic nature of the CMBR data in the 
$W1$ band seen in the 3-year and 5-year data is absent in the current 9-year data. 

\section{Acknowledgements:}Some of the results in this paper have been derived using the Healpix package \cite{gorski05}. We are
grateful to Pankaj Jain for a very useful discussions. Finally we
acknowledge the use of Planck data available from NASA LAMBDA site (http://lambda.gsfc.nasa.gov). 


\begin{thebibliography}{99}

\bibitem{Birch1982} Birch P., Nature, 298, p.451, (1982)

 \bibitem{Jain1999}
P. Jain and J. Ralston, Modern Physics Letters A, 14,  417  (1999).

\bibitem{Hutsemekers1998}
D. Hutsemekers, Astronomy \& Astrophysics, 332,  410  (1998).

\bibitem{Ralston2004}
J. Ralston and P. Jain, International Journal of Modern Physics D 13,
  1857  (2004).

\bibitem{Schwarz2004}
D. Schwarz, G. Starkman, D. Huterer, and C. Copi, Physical Review Letters, 93,  221301  (2004).

\bibitem{Samal2008} Samal P. K., Saha R., Jain P. and Ralston J. P., MNRAS, MNRAS, 385, 4, p.1718 (2008) 

\bibitem{Samal2009} Samal P. K., Saha R., Jain P. and Ralston J. P., MNRAS, 396, 1, 511–522 (2009) 

\bibitem{Aluri2011} Aluri Pavan K., Samal P. K.,  Jain Pankaj, Ralston John  P., MNRAS, 414, 1032–1046 (2011)

\bibitem{Copi2007} Copi C. J., Huterer D., Schwarz D. J. and Starkman G. D., Phys. Rev. D, 75, 2, 023507 (2007) 

\bibitem{Costa2004} de Oliveira-Costa A., Tegmark M., Zaldarriaga M. and Hamilton A., Phys. Rev. D, 69, 6, 063516 (2004) 

\bibitem{Copi2004} Copi C. J., Huterer D. and Starkman G. D., Phys. Rev. D, 70, 4, 043515 (2004) 

\bibitem{hinshwa2007} Hinshaw, G. {\it et al}, ApJ Suppl 170, 288 (2007) 

\bibitem{gorski05} Gorski K. M. {\it et al}, ApJ, 622, 759 (2005)

\bibitem{Bennett2013} Bennett, C.L., {\it et.al}, ApJS., 208, 20B (2013)

\bibitem{Hinshaw2007} G. Hinshaw, {\it et al}, ApJS, 170, 288 (2007)

\bibitem{Planck2014b} Planck Collaboration A\&A 571, A1 (2014)

\end{thebibliography}
\end{document}